\documentclass{aa}
\usepackage{times}
\usepackage{epsfig,float}
\usepackage{graphics}
\usepackage{amssymb}

\begin{document}

\title{Constraints on   the  optical afterglow emission of   the short/hard
  burst  GRB~010119\thanks{Based on   observations   made with the   Nordic
    Optical Telescope,  operated  on the  island of   La Palma  jointly  by
    Denmark,   Finland,  Iceland,  Norway,   and  Sweden,  in  the  Spanish
    Observatorio  del   Roque de  los    Muchachos  of  the   Instituto  de
    Astrof\'{\i}sica de Canarias.}}

\titlerunning{The short/hard burst GRB~010119}

\author{
   J. Gorosabel \inst{1}
   \and M.I. Andersen \inst{2}
   \and J. Hjorth \inst{3}
   \and H. Pedersen \inst{3}
   \and B.L. Jensen \inst{3}
   \and J.U. Fynbo \inst{4}
   \and H.J.  Lehto \inst{5,6}
   \and S.  Katajainen \inst{5}
   \and K. Hurley \inst{7}
   \and N. Lund   \inst{1}
}

\offprints{J. Gorosabel}

\institute{ 
           Danish Space Research Institute,
           Juliane Maries Vej 30, DK--2100 Copenhagen \O, Denmark;
           jgu@dsri.dk, nl@dsri.dk
           \and
           Division of Astronomy, University of Oulu P.O. Box 3000,
           FIN-90014 University of Oulu, Finland;
            michael.andersen@oulu.fi
            \and
           Astronomical Observatory, 
           University of Copenhagen, 
           Juliane Maries Vej 30, DK--2100 Copenhagen \O, Denmark;
           jens@astro.ku.dk,holger@astro.ku.dk,brian\_j@astro.ku.dk
           \and
           European Southern Observatory,
           Karl--Schwarzschild--Stra\ss e 2,
           D--85748 Garching, Germany;
           jfynbo@eso.org
           \and
           Tuorla Observatory, V\"ais\"al\"antie 20, 
           FI-21500 Piikki\"o, Finland; harry.lehto@astro.utu.fi, seppo.katajainen@astro.utu.fi
           \and
           Department of Physics, University of Turku, FI-20140, Finland.
           \and
            University of California, Berkeley,
            Space Sciences Laboratory,
            Berkeley, CA 94720--7450;
            khurley@ssl.berkeley.edu
           }

\mail{\tt jgu@dsri.dk}

\date{Received  / Accepted }

\abstract{We report optical observations of the short/hard burst GRB~010119
  error  box,  one of  the   smallest  error  boxes  reported  to date  for
  short/hard   GRBs.  Limits  of R~$>22.3$  and  I~$>21.2$  are imposed  by
  observations carried  out $20.31$ and $20.58$   hours after the gamma-ray
  event, respectively.  They represent the most constraining limits imposed
  to date on   the  optical emission  from  a  short/hard  gamma-ray  burst
  afterglow.   }

\maketitle


\section{Introduction}

The bimodal   distribution   (Hurley, \cite{Hurl92}; Kouveliotou  et   al.  
\cite{Kouv93})   of  gamma-ray bursts (GRBs)   separating  them  into short
duration ($T<2\rm\, s$) and long duration ($T>2\rm\, s$) bursts was already
known  before the era  of Beppo-SAX.  The short bursts  tend to have harder
spectra than the long bursts (Kouveliotou et al.  \cite{Kouv93}; Dezalay et
al.\ \cite{Deza96}).   Hence, in the present paper  the short and long GRBs
will   be  hereafter  named  as ``short/hard''   and  ``long/soft'' bursts,
respectively.  Several years later, an analysis  of the Third BATSE Catalog
indicated that, in addition to these two classes of bursts, there may exist
a third, intermediate soft-spectrum class of GRBs with durations $2$s $ < T
< 5$s (Mukherjee et al. \cite{Mukh98}).

Although the existence of this third intermediate duration class of GRBs is
still under  debate, in this  paper we  have considered  them as a separate
class. Thus, we consider a tri-modal distribution of GRBs; short/hard ($T <
2$s), intermediate ($2$s $ < T < 5$s) and long/soft ($T > 5$s) bursts.

In the  simplest scenario, the  short/hard  bursts may  be explained by the
merging of  the two  compact components  of  a binary system  (Lattimer  \&
Schramm \cite{Latt74}, Eichler et  al.  \cite{Eich89}), although other more
exotic theories  like the evaporation  of primordial black-holes could also
explain the  observed  properties of   short/hard bursts (Page  \&  Hawking
\cite{Page76}; Cline,  Matthey, \& Otwinowski \cite{Clin99}).  According to
the merging model, GRBs would occur  in very-low density environments, with
very faint  or even  no  afterglows  at all.   The  observed properties  of
long/soft bursts can be better accommodated in the context of the collapsar
model (MacFadyen \& Woosley \cite{MacF99}).

Castro-Tirado  et al.  (\cite{Cast01})  have recently reported the possible
detection of  the  prompt optical flash  4 min  after the  short/hard burst
GRB~000313, suggesting that  short/hard   GRBs only show   optical emission
shortly after the  gamma-ray event, with no  afterglows at all.   This fact
would favour the  models that relate short/hard GRBs  to  binary mergers in
low-density environments.

All  the GRBs for   which optical, X-ray,  and  radio afterglows have  been
discovered to date belong to the long/soft GRB class, with the exception of
a  couple of potential intermediate  duration GRBs (GRB  000301C, Jensen et
al.  \cite{Jens01};   GRB  991014,  in't  Zand   et  al.  \cite{IntZ00}).   
GRB~991106  was  preliminarily  classified as   a   possible short/hard GRB
(Gandolfi et al., \cite{Gand99a}),  a more detailed  analysis noted that it
belongs  to    the  long/soft or  intermediate   class   (Gandolfi  et al.,
\cite{Gand99b}).     Hereafter the    optical  upper  limits   reported for
GRB~991106 (Castro-Tirado      et al.  \cite{Cast99a},    Williams et   al. 
\cite{Will99},   Jensen     et  al.   \cite{Jens99},   Gorosabel    et  al. 
\cite{Goro99})  will  not be considered as   constraints on short/hard GRB
afterglows.  Thus, no afterglows have been  detected to date for short/hard
GRBs, so their origin as well as their distance  scale remain unknown.  The
detection of a  short/hard optical afterglow similar to  the  ones seen for
long/soft   bursts   would argue    against the   low-density   environment
hypothesis.

Upper limits for  the prompt optical  emission  (response times  $<$ 5 min,
R$<15$), as well as for  the afterglow emission (response  times $>$ 5 min,
R$<16$) have been reported by Kehoe et al.  (\cite{Keho01}).  Hurley et al.
(\cite{Hurl01a}) has   recently    reported improved   positions  of   four
short/hard GRBs  determined by the Interplanetary  Network (IPN) as well as
several   constraining upper limits on their    afterglow optical and radio
emission.

Fynbo et al.  (\cite{Fynb01}) have recently argued that $\sim\!75\%$ of the
upper limits reported to date for  long/soft GRBs are compatible with faint
afterglows as the one of GRB~000630, which are unreachable with most of the
current long response times and shallow detection limits.  Reichart \& Yost
(\cite{Reic01a}) suggest that the  majority of rapidly, well-localized GRBs
with undetected optical afterglows are most likely the result of extinction
by dust  in  the circumburst  medium.   This idea is  supported by Lazzati,
Covino \& Ghisellini (\cite{Lazz00}) who claim that  the low detection rate
can not be explained by adverse observing conditions or delay in performing
the observations.   Thus, the existence of  intrinsically dark bursts would
imply that  the UV flash  and the X-ray afterglow  do not  destroy the dust
responsible   of   the   optical  extinction   (Reichart  \cite{Reic01b}).  
Panaitescu, Kumar  \&  Narayan (\cite{Pana01}) predict for  short/hard GRBs
faint optical afterglows, exhibiting  typically  R$\gtrsim 23$ a few  hours
after the gamma-ray event.

In  the absence of   optical afterglow detections  from short/hard  bursts,
deeper  and earlier upper limits  on the afterglow flux  is the only way to
answer one of the main  open questions regarding  short/hard GRBs; do  they
exhibit  optical   afterglows?   In  the   present  paper  we  discuss this
questions, reporting   constraining   R and  I-band   upper  limits  on the
afterglow  optical   emission  from  GRB~010119.   In  Sec.~\ref{loca}  the
high-energy  properties  as well  as  the  localisation  of  GRB~010119 are
reported.  Sec.~\ref{obs} describes  the  optical observations obtained  at
the Nordic Optical Telescope  (NOT).  In Sec.~\ref{discus} our upper limits
are compared  to other constraints given  in  the literature for long/soft,
intermediate,   and    especially  short/hard  GRB  afterglows.    Finally,
Sec.~\ref{conclusion} summarizes the conclusions of our study.

\begin{figure}[t]
\begin{center}
\epsfig{file=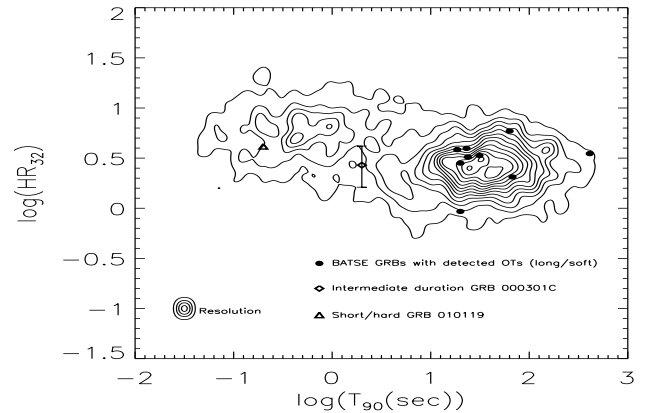, width=9.4cm , height=6cm }
\caption{\label{contour} 
  A           contour       plot     showing        the   duration-hardness
  ($log(T_{90})$-$log(H_{32})$) distribution of  BATSE bursts.   The filled
  circles  represent the 9  BATSE  bursts with identified  counterparts for
  which data on fluence and duration are  available.  All of them belong to
  the long/soft class of  GRBs.  The open diamond located  in the centre of
  the figure represents the intermediate duration burst GRB 000301C (Jensen
  et  al.   \cite{Jens01}).   The open  triangle to  the  left of  the plot
  represents  GRB~010119.  As   can be   seen, GRB~010119  belongs  to  the
  short/hard class of GRBs.  Errors in the  BATSE data are smaller than the
  symbol  size.  Contour levels scale linearly.   The centroid in the lower
  left corner indicates the resolution.}
\end{center}
\end{figure}

\begin{table*}[t]
\begin{center}
\caption{Journal  of the \object{GRB~010119} optical observations}

\begin{tabular}{lcccccccr}
Telescope&Date (UT)&Seeing&Airmass&Effective&filter&Exp.\ time&Lim. Mag.$^{\dag\dag}$ \\
         &         &arcsec&       &Airmass  &      &(sec)      &(3$\sigma$)\\
\hline
NOT&20.2703--20.2829/01/2001&$1.1\times1.8^{\star}$&$5.34-3.84$&4.45&R&3$\times$300&22.3\\
NOT&20.2840--20.2920/01/2001&$1.2\times1.5^{\star}$&$3.84-3.25$&3.57&I&2$\times$300&21.2\\
NOT&21.2689--21.2831/01/2001&$3.0^{\dag}$          &$5.15-3.59$&4.27&R&3$\times$300&20.9\\
NOT&21.2848--21.2977/01/2001&$1.7^{\dag}$          &$3.59-3.05$&3.36&I&3$\times$300&20.2\\
NOT&29.2035--29.2257/05/2001&$0.75$                &$1.10-1.13$&1.11&R&2$\times$900&24.5\\
NOT&14.0390--14.0261/08/2001&$1.0$                 &$1.19-1.23$&1.21&I&3$\times$300&23.0\\
\hline
\multicolumn{8}{l}{$\star$ Elongated point spread function due to the differential chromatic refraction of the atmosphere.}\\
\multicolumn{8}{l}{$\dag$  Through clouds.}\\
\multicolumn{8}{l}{$\dag\dag$  Not corrected for Galactic extinction.}\\
\hline 
\label{table1}
\end{tabular}
\end{center}
\end{table*}

\begin{figure*}[t]
\begin{center}
\epsfig{file=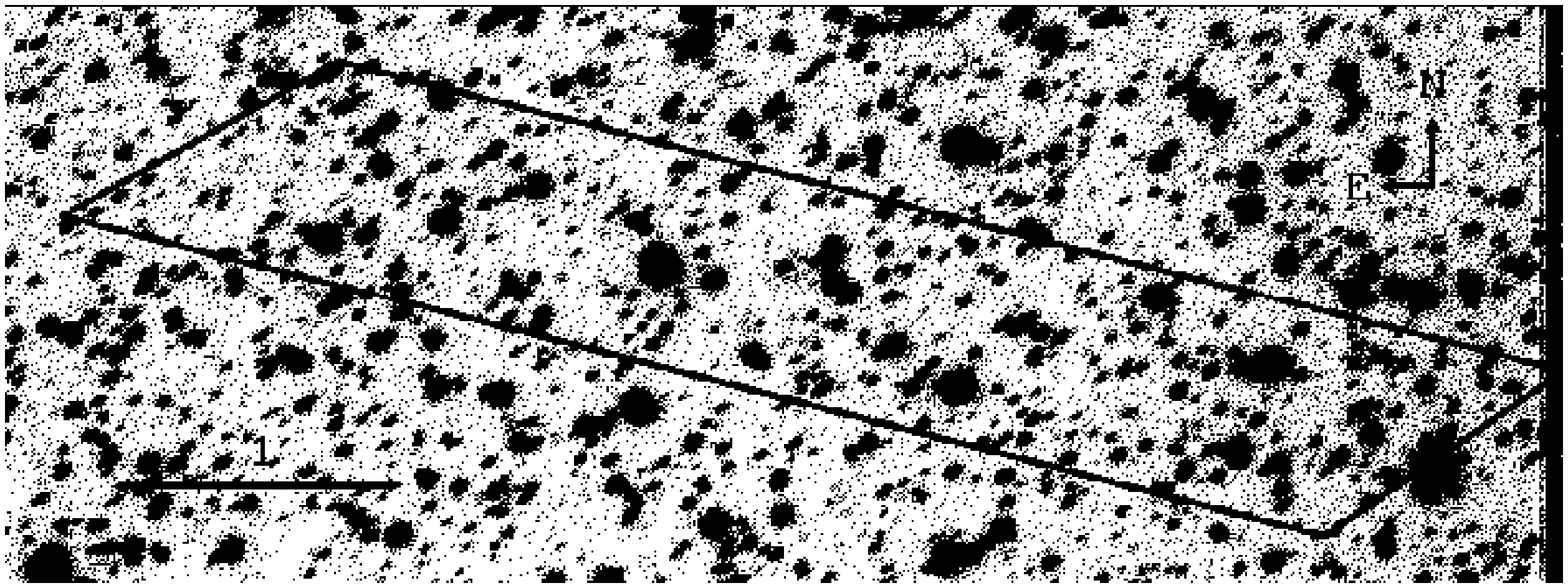, width=6.4cm,angle=-90}
\epsfig{file=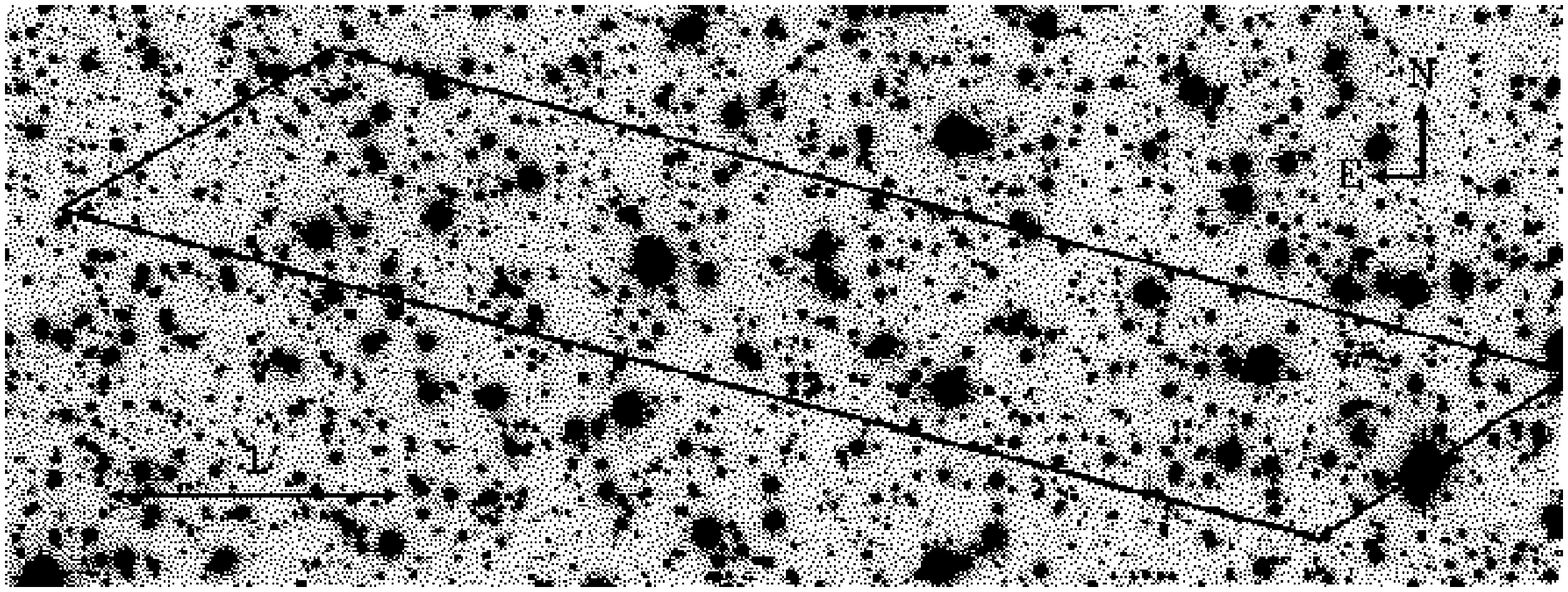,width=6.4cm,angle=-90}
\caption{\label{images} {\it  Upper  panel:} The plot shows the result
  of   co-adding the  three    R-band exposures taken   at  NOT  on January
  $20.2703$--$20.2829$ UT 2001.  We have overplotted  the refined IPN error
  box  given  by Hurley  et    al.  (\cite{Hurl01a}).   The  stars show  an
  elongated profile due to the  very high airmass  of  the image ($5.34  <$
  airmass $< 3.84$).  However, the image was deep enough to impose a severe
  constraint on the optical  emission of the  burst (mean star profile FWHM
  $\sim1\farcs45$, limiting magnitude  R$=22.3$).   {\it Lower panel:}  The
  figure displays the deep comparison  image (limiting magnitude  R$=24.5$)
  taken on  May  $29.2035$--$29.2257$ UT 2001.    No  variable sources were
  found.}
\end{center}
\end{figure*}

\begin{figure*}[t]
\begin{center}%
\epsfig{file=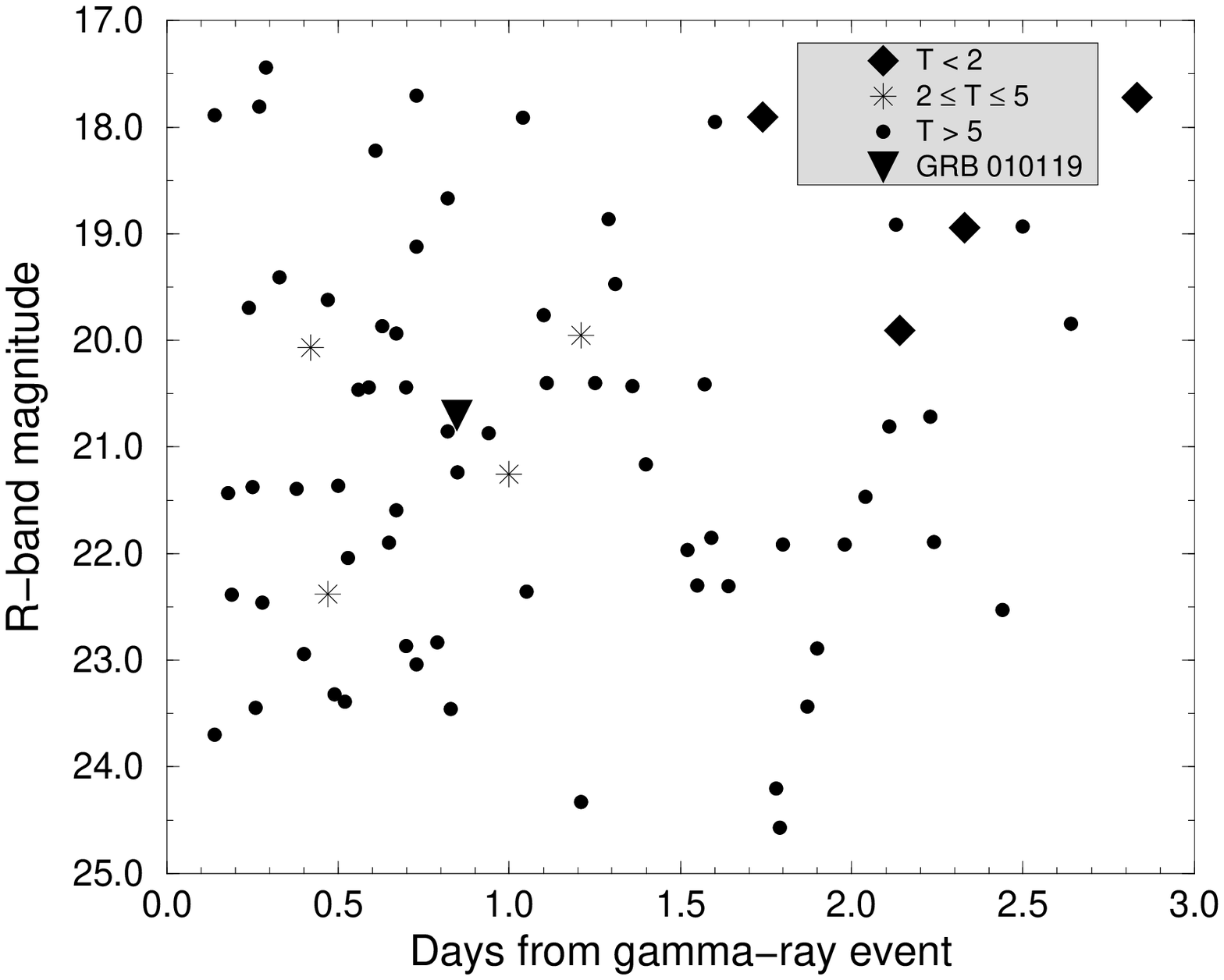, width=8.95cm}
\epsfig{file=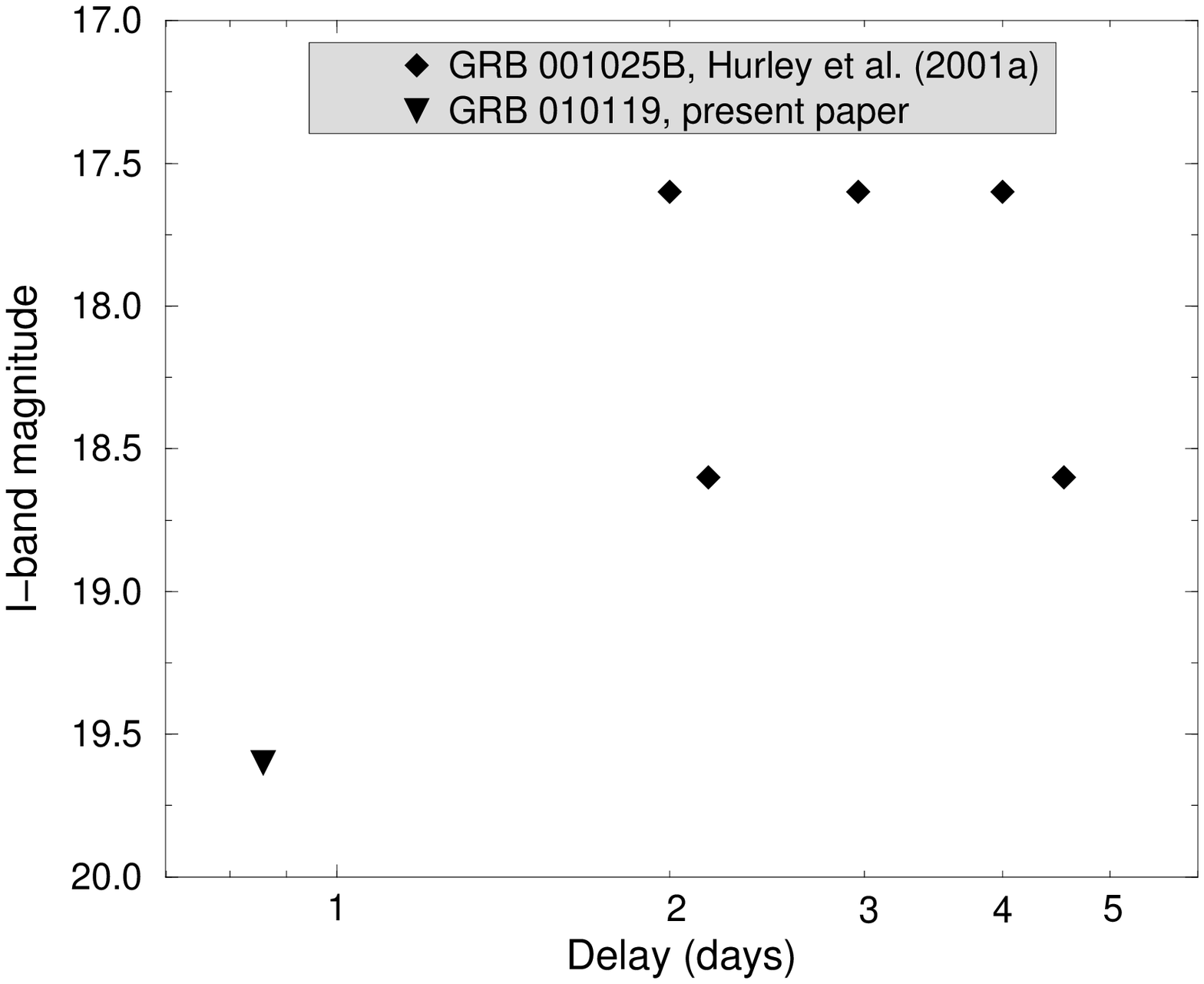, width=8.95cm}
\caption{\label{limits} {\em left panel:} 
  The  figure   shows  several  R-band  upper limits   for  long/soft burst
  (circles), intermediate bursts   (stars) and short/hard burst  afterglows
  (diamonds).  The  upper limit imposed  $20.31$  hours after GRB~010119 is
  represented by a triangle.  As can be seen, the constraint imposed by the
  triangle  is comparable to  the  reported  upper-limits of long/soft  and
  intermediate  GRB afterglows,  and  is  much more  constraining than  the
  R-band upper limits reported to date for other short/hard GRB afterglows.
  {\em right panel:} The plot displays  the I-band upper limits reported by
  Hurley et al.   (2001a) for GRB~001025B  (diamonds) as well as the I-band
  upper limit imposed in this paper for GRB~010119, $20.58$ hours after the
  burst  (triangle).  As can  be seen, also in the   I-band the triangle is
  more  constraining than  the limits  reported to date  for short/hard GRB
  afterglows.}
\end{center}
\end{figure*}

\section{Localisation of the GRB}
\label{loca}
GRB~010119 was detected by the IPN, composed by the Ulysses, NEAR, WIND and
Beppo-SAX spacecraft,  on January  $19.430306$  UT  2001  (Hurley et  al.   
\cite{Hurl01a}).   As  observed   by Ulysses,    it   had  a duration    of
approximately $0.2$ seconds  (Hurley et al.   \cite{Hurl01b}).  Integrating
the analytic  expression of the spectral  fit of GRB~010119 given by Hurley
et al.  (\cite{Hurl01a}) the  hardness-ratio between the BATSE channels $2$
and  $3$ (H$_{32}$;   the 100-300 keV   fluence  divided by the 50-100  keV
fluence) H$_{32} = 4.14$ is obtained.

In Fig.~\ref{contour} the duration  distribution of $2115$ BATSE GRBs, from
the revised Fourth BATSE GRB Catalogue  (Paciesas et al.  \cite{Paci99} and
the      Current        BATSE       GRB        Catalogue         at    {\tt
  http:\-//www.\-batse.\-msfc.\-nasa.\-gov/\-batse/\-grb/\-catalog/\-current})
is   displayed.   The bi-modal   distribution   composed by  long/soft  and
short/hard  classes  of bursts can  easily  be seen.  Additionally, we have
overplotted 9 long/soft BATSE  bursts with identified afterglows for  which
data  on fluence and duration are  available (filled circles).  The diamond
represents the short-intermediate GRB~000301C (Jensen et al. \cite{Jens01})
and the triangle  shows GRB~010119.  As can be  seen, GRB~010119 belongs to
the short/hard class of gamma-ray bursts.

The coordinates of the centre of the improved GRB~010119  IPN error box are
$\alpha_{2000}=18^{h}    53^{m}     46.17^{s}$,   $\delta_{2000}=11^{\circ}
59^{\prime} 47\farcs04$  (Hurley et  al.~\cite{Hurl01a}).  The size  of the
error box is $3.3$ arcmin$^2$,  significantly smaller than the  preliminary
$11.0$ arcmin$^2$  box first reported by  Hurley  et al.  (\cite{Hurl01b}). 
In fact, the error box of GRB~010119 is one of the smallest IPN error boxes
reported so  far for short/hard   GRBs.  This fact,   as well as the  early
dissemination  of the IPN  position  ($14.7$ hr, the earliest dissemination
among the 4 short/hard  bursts reported by Hurley  et al.  \cite{Hurl01a}),
enabled us to obtain early  data, providing an exceptional opportunity  for
detecting the counterpart of a short/hard burst.
 
\section{Observations}
\label{obs}

The optical observations reported in   the present paper were carried   out
from the NOT, equipped with   the Andaluc\'{\i}a Faint Object  Spectrograph
and  Camera (ALFOSC).  The   field of  view of   ALFOSC ($6\farcm 5  \times
6\farcm 5$) allowed to cover both the entire initial IPN error box, as well
as the  refined  one.  The observations  started  $20.16$ hours   after the
gamma-ray event at an airmass of $5.34$.  The good  transparency of the sky
as well as the excellent seeing conditions allowed us to obtain images with
a  Full Width Half Maximum (FWHM)  star profile of $1\farcs1\times1\farcs8$
at  this extreme airmass.  The  stars appear elongated perpendicular to the
horizon  due to the atmospheric  differential chromatic refraction.  On May
$20$ and $21$ (the  first  two nights  of observations)  the Sun was   only
$\sim$$37$ degrees from  the GRB field,  so on these dates the observations
only lasted $31.25$ (May  $20.2703$--$202920$ UT) and $41.47$  minutes (May
$21.2689$--$21.2977$) respectively,   immediately  before dawn  (see  Table
\ref{table1}).

Due to the high airmass  and the close  proximity to the Sun the background
varies   substantially from  image to   image.  In  order   to optimize the
combination of the  data acquired  through a  given filter, the  individual
images have been weighted  with the inverse  of their mode.  As the airmass
gradient was very high during the exposures, it  is convenient to calculate
the  effective  airmass of  the  resultant  co-added  image.  The effective
airmass  was calculated weighting  the  mean   airmass of  each  individual
exposure with the same weights  used for the   combination of the  co-added
image.

Another  difficulty was  the low Galactic   latitude ($b=4.93^{\circ}$) and
very crowded field of GRB~010119.  The corresponding Galactic extinction in
the  R and   I-bands  are  A$_{\rm   R}=1.59$  and A$_{\rm   I}=1.15$  mag,
respectively (Schlegel, Finkbeiner \& Davis \cite{Schl98}).

The comparison  of the images was  carried  out with SExtractor  (Bertin \&
Arnouts \cite{Bert96}), which enabled us to  deblend the overlapping stars. 
The co-added images were WCS calibrated and the magnitudes of the spatially
coincident sources were derived.  No source above the upper limits given in
Table~\ref{table1} exhibited magnitude differences $\Delta \!  m > 0.2$~mag
in either the R or I-bands. Fig.  \ref{images} shows  the refined error box
overplotted on the   R-band  images  taken at    the NOT on  2001   January
$20.2703$--$20.2829$ UT and   2001 May $29.2035$--$29.2257$ UT (see   Table
\ref{table1}).

The  calibration was performed in  August  2001 observing the Landolt field
SA113 at   a similar airmass  as the  field  (Landolt \cite{Land92}), which
allowed us to obtain the R and I-band magnitudes of several secondary stars
in  the error box.  The  errors  in the calibration,   and in the 3$\sigma$
upper limits displayed in Table~\ref{table1},  are smaller than $\Delta \!$
R~$\lesssim0.05$ mag  and $\Delta \!$   I~$\lesssim0.04$ mag, respectively. 
These   errors  are  of  no  significance  for   the   results discussed in
Sec.~\ref{discus}.
 
\begin{figure*}[t]
\begin{center}
\epsfig{file=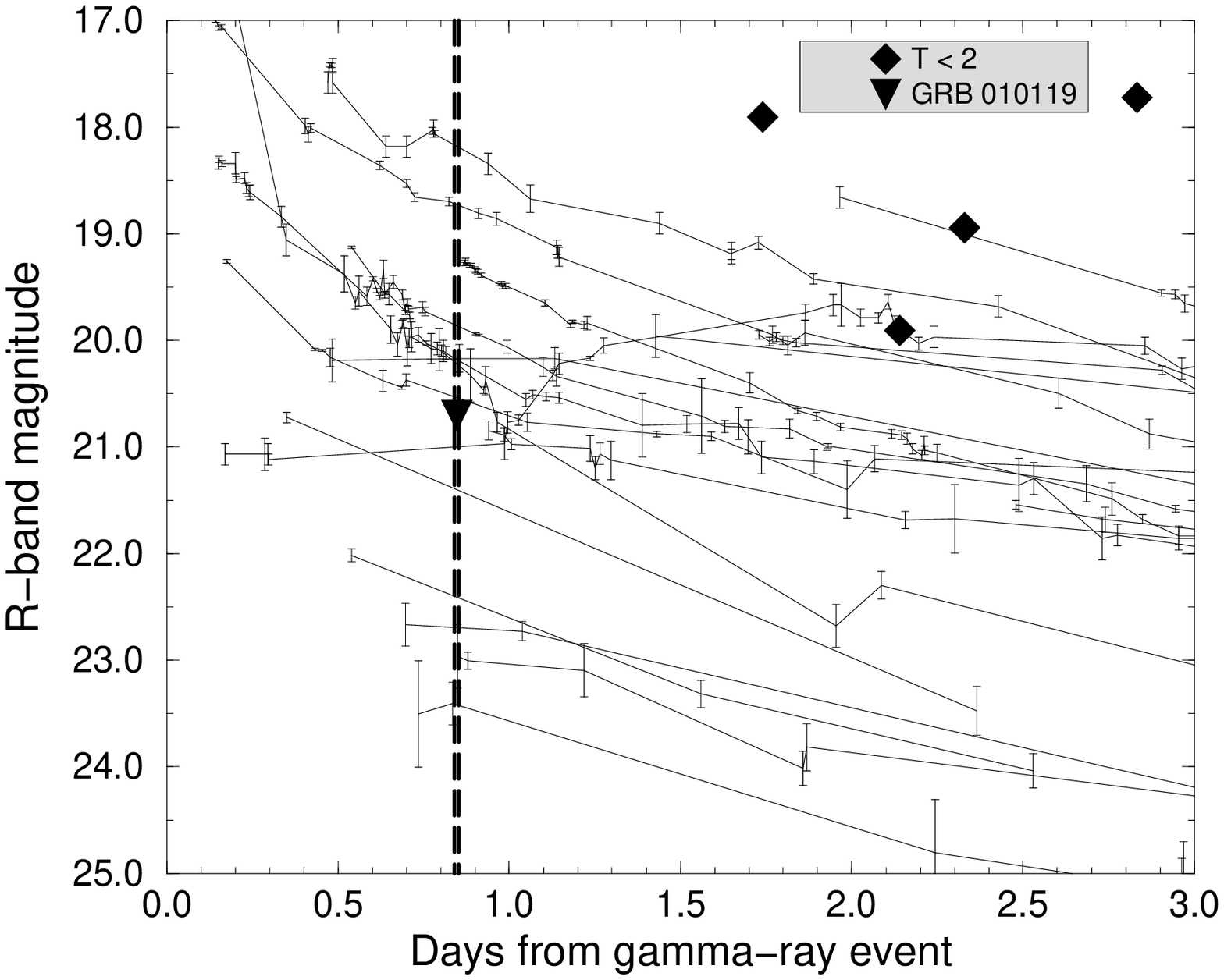, width=8.95cm}
\epsfig{file=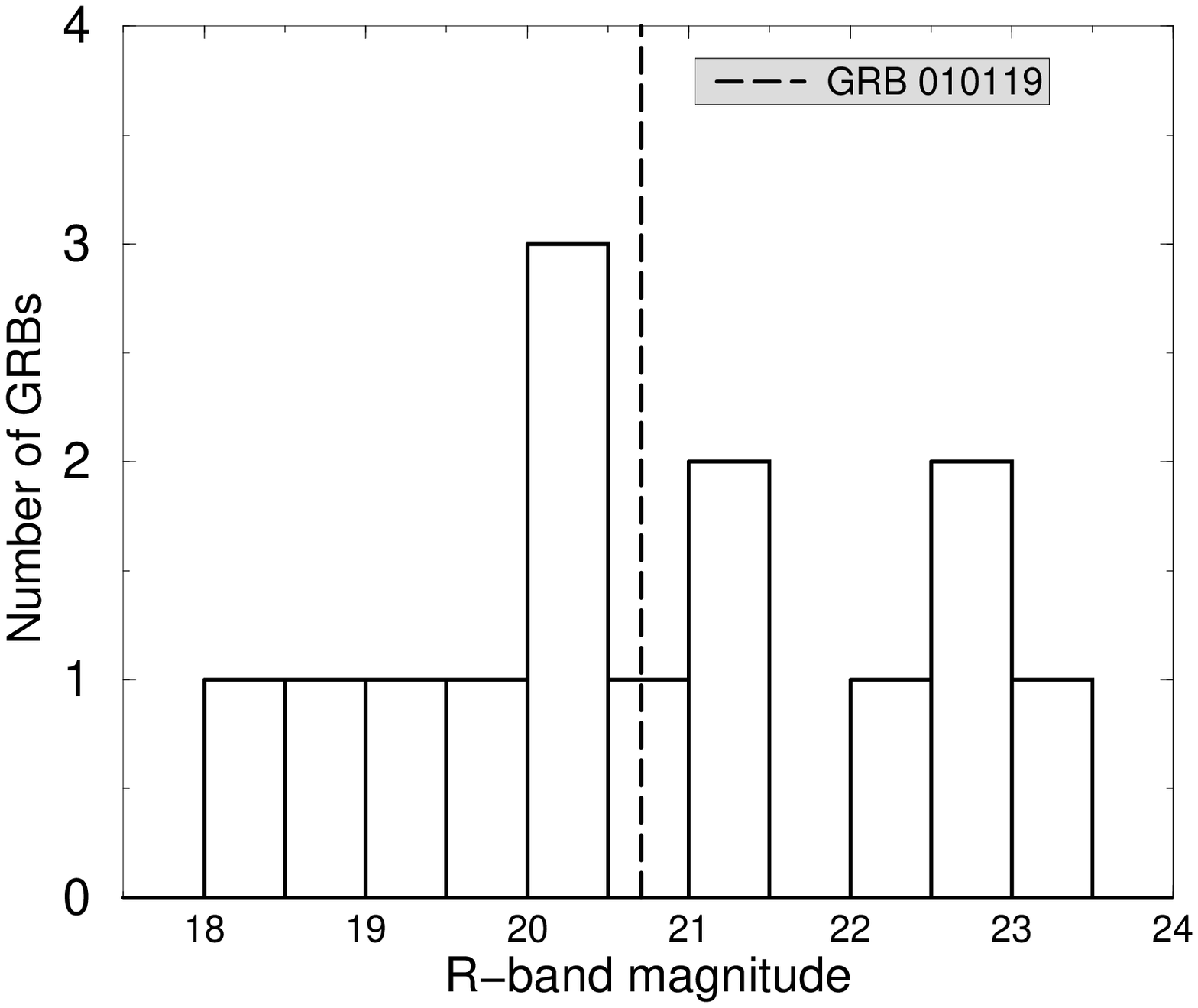, width=8.95cm}
\caption{\label{lightcurves} {\em left panel:} 
  The lightcurves of  20 optical afterglows  for which there are detections
  in the R-band.   As in the left panel  of Fig.~\ref{limits} the  diamonds
  represent the R-band upper limits reported to date for afterglow emission
  from short/hard GRBs.  The triangle shows the  R-band upper limit imposed
  for GRB~010119.  The vertical dashed band at the  epoch of the GRB~010119
  R-band  upper limit  intersects 14 of  the  20 lightcurves.  Among the 14
  lightcurves,  8 intersect the vertical  line above the  triangle.  On the
  other hand, only one of the four diamonds (the one representing the upper
  limit  given by  Price  et al.  \cite{Pric01})  is  in  the detectability
  region, being below 3 out of 20 lightcurves.  {\em right panel:} The plot
  shows the distribution  of the magnitudes  of the 14 lightcurves from the
  left panel at the epoch of the GRB~010119 upper limit, represented by the
  dashed line.}
\end{center}
\end{figure*}

\section{Discussion}
\label{discus}

We have compiled from the literature the R-band  upper limits for long/soft
($T >5$s), intermediate ($2$s~$ <  T <5$s) and short/hard  ($T < 2$s)  GRBs
for which no afterglow was found and plotted them in the left panel of Fig.
\ref{limits}.   In  order   to compare  our    measurements  only  to  deep
observations (R~$>17$) we have not covered the  region defined by the upper
limits of short/hard burst afterglows given by robotic telescopes (Kehoe et
al.  \cite{Keho01}).    The  upper   limits  have been    collected  mainly
considering Table~2 of Fynbo et al.   (\cite{Fynb01}), Table~3 of Hurley et
al.   (\cite{Hurl01a})  and  extending them   by using GCN  notices  up  to
GRB~010707.  In the right  panel of Fig.~\ref{limits} the best  I-band
upper limits reported to date for short/hard  bursts (see Table~3 of Hurley
et   al.  \cite{Hurl01a})  are  compared to  the  I-band  limit imposed for
GRB~010119 in this study.  All  magnitudes shown if Figs.  \ref{limits} and
\ref{lightcurves},   as well as   those  discussed  in the   following, are
corrected for their corresponding Galactic extinction (Schlegel, Finkbeiner
\& Davis \cite{Schl98}).

The  R-band   upper limit presented  in this   paper  (see triangle  of the
Fig.~\ref{limits} left   panel) is much more   constraining than  the other
upper limits reported to date  for afterglow emission from short/hard  GRBs
(diamonds).  In fact, the R-band observation was done earlier and is deeper
than any of  the $7$ R-band  upper limits previously  reported by Hurley et
al.  (\cite{Hurl01a}).

In order to compare our R-band upper  limits to the  ones reported by other
authors for short/hard GRB afterglows,  all of them  have to be shifted  to
the same  epoch  assuming a power law   decay $F_{\nu} \sim t^{-\alpha_{\rm
    R}}$ with a  given  value of  the decay index  $\alpha_{\rm R}$.    The
closest  diamond  from the  triangle   in Fig.    \ref{limits} left   panel
represents the  upper limit reported by  Price  et al.  (\cite{Pric01}) for
GRB~010119.  This upper limit is $0.8$ mag  shallower than the triangle and
was imposed  $\sim$$51$ hours after the gamma-ray  burst.  If we assume the
conservative  case of a shallow  afterglow decay with $\alpha_{\rm R}=1.0$,
the measurement by Price et  al.   (\cite{Pric01}), would correspond to  an
observed  magnitude of   R=$20.5$ at  the  time  of our  measurement, which
corresponds to  a ratio of  $5.2$  between both  flux sensitivities.   This
relative comparison  between sensitivities has to  be considered as a lower
limit because, in  the more realistic case, $\alpha_{\rm  R}  > 1$ for  GRB
optical afterglows.

The most constraining I-band upper limit  reported to date for a short/hard
GRB  afterglow corresponds to the observation  carried out  with the $40$''
Telescope of Las  Campanas  for GRB~001025B (Hurley  et al.~\cite{Hurl01a};
I~$>21.5$,   $52$ hours after  the  GRB.  See the    closest diamond to the
triangle of the Fig.~\ref{limits}  right panel).   The  delay and  the high
extinction of GRB~001025B  (A$_{\rm  I}=2.90$ mag; Schlegel,  Finkbeiner \&
Davis \cite{Schl98}) makes  this upper  limit  less constraining than   the
I-band  observation carried  out  on January   $20.2840$--$20.2920$  UT for
GRB~010119  (triangle of right   panel of  Fig.~\ref{limits}).  Assuming  a
shallow   afterglow decay   of   $\alpha_{\rm    I}=1.0$,  we  derive     a
contemporaneous flux sensitivity ratio of  $9.6$ between both upper limits,
once the relative Galactic reddening factor  is introduced.  Therefore, the
I-band  upper limit imposed  on January $20.2840$--$20.2920$ UT (I~$>21.2$,
$20.58$  hours after  the    burst, see Table~\ref{table1})  is  the   most
constraining upper  limit imposed to date in  this filter  for a short/hard
GRB afterglow.

From the left panel  of Fig.  \ref{lightcurves} it  is evident that the all
previous detection limits  on  short/hard afterglows (diamonds)  would have
missed  most of the  afterglows (only  one  diamond is consistent  with the
detectability region of the long/soft GRB  afterglows, being below 3 out of
20 lightcurves, so the success ratio would be $\sim  15\%$), if we make the
ad-hoc  assumption that short/hard  bursts  have afterglow  characteristics
similar to those  of long/soft bursts.  The R-band  limit  on GRB~010119 is
sufficiently  deep that an about $\sim$$60$\%  of  the long/soft afterglows
with  an  R-band  detection  within a  day  of the   burst  would have been
detected, as illustrated in the right panel of Fig.  \ref{lightcurves}.

However, three  out of   four  long bursts  are never  detected  at optical
wavelengths.  Taking this into account, the afterglow detection probability
for GRB~010119  would be $\sim$$15$\%.  Some  undetected bursts  are likely
dark  due to late  follow-up,  suggesting that a  more realistic  detection
probability  with the  limit achieved for   GRB~010119 is of  the order  of
$\sim$$25$\%. Independently of the fraction of intrinsically dark GRBs, the
detection probability of our  promptest R-band observation is $\sim4$ times
higher  (8/14  vs 3/20) than  previous   R-band  measurements reported  for
short/hard GRB    afterglows, assuming  that    short/hard and    long/soft
afterglows have similar characteristics.

\section{Conclusion}
\label{conclusion}

The R-band and  I-band limits imposed $20.31$  and $20.58$ hours  after the
gamma-ray  event represent the   most constraining measurements reported to
date on the optical afterglow emission from a short/hard burst. 

If  GRB~010119 had   shown an optical   evolution  similar  to the  typical
long/soft and  intermediate  duration optical afterglows,  our observations
would  have  had   a probability of   $\sim$$60$\%   to detect   its R-band
counterpart.  Assuming the conservative case  that only $\sim$$25$\% of the
long/soft bursts exhibit optical emission, a lower limit of $\sim$$15$\% is
derived for the success ratio of our  R-band upper limit.  


Therefore, our   observations    are compatible  with  a   completely  dark
short/hard  GRB afterglow     (as suggested  by  Castro-Tirado  et    al.   
\cite{Cast01})  or with a  long/soft-like burst with faint optical emission
as GRB~000630 (Fynbo   et al.  \cite{Fynb01}).   Our upper  limits are also
consistent  with the  predictions   given by  Panaitescu, Kumar \&  Narayan
(\cite{Pana01}) for short/hard GRB optical afterglows.

A large number of constraining  upper limits would  be necessary to clarify
whether short/hard  bursts  do not exhibit afterglows,  as  expected in the
context of the low-density environments.

\section{Acknowledgments}

JG acknowledges  the receipt  of a   Marie  Curie Research Grant   from the
European Commission.  MA acknowledges the support of the University of Oulu
astrophysics group. JH acknowledges support from the Danish Natural Science
Research Council (SNF).  SK wishes to thank the  Finnish Academy of Science
and Letters (Academia Scientiarum Fennica) for support.  The work of HJL is
partially funded by Finnish Academy grants number  71355 and 44011. Support
for the  Ulysses GRB experiment is  provided by JPL  Contract 958056.  NEAR
data  analysis was supported under  NASA Grants NAG  5-3500 and NAG 5-9503. 
We also   thank Scott Barthelmy for  developing  and maintaining   the GCN,
without which most counterpart searches  could not be made.  This  research
has made use of the NASA/IPC Extragalactic Database (NED) which is operated
by the Jet Propulsion Laboratory, California Institute of Technology, under
contract with National Aeronautics and Space Administration.  We appreciate
the support of  the NOT  staff astronomers.   The data presented  here have
been taken using ALFOSC, which is owned by  the Instituto de Astrofisica de
Andalucia   (IAA)  and  operated at  the   Nordic Optical   Telescope under
agreement between IAA and the  NBIfAFG  of the Astronomical Observatory  of
Copenhagen.

\end{document}